\documentclass[a4paper,11pt]{article}
\usepackage{jheppub}

\usepackage{graphicx,placeins}
\usepackage{wrapfig}
\usepackage[utf8]{inputenc}

\newcommand\beq{ \begin{eqnarray} }
\newcommand\eeq{ \end{eqnarray} }

\usepackage{color}

\title{\boldmath Two-colour QCD phases and the topology at low temperature and high density}

\author[a]{Kei Iida,}
\author[a,b,c]{Etsuko Itou} 
\author[a]{and Tong-Gyu Lee}%\note{}

\affiliation[a]{Department of Mathematics and Physics, Kochi University, Kochi 780-8520, Japan}
\affiliation[b]{Department of Physics, and Research and Education Center for Natural Sciences, Keio University, 4-1-1 Hiyoshi, Yokohama, Kanagawa 223-8521, Japan}
\affiliation[c]{Research Center for Nuclear Physics (RCNP), Osaka University, Osaka 567-0047, Japan}

\emailAdd{iida@kochi-u.ac.jp}
\emailAdd{itou@yukawa.kyoto-u.ac.jp}
\emailAdd{tonggyu.lee@yukawa.kyoto-u.ac.jp}

\abstract{%
We delineate equilibrium phase structure and topological charge distribution 
of dense two-colour QCD at low temperature by using a lattice simulation 
with two-flavour Wilson fermions that has a chemical potential $\mu$ and 
a diquark source $j$ incorporated.  We systematically measure the diquark 
condensate, the Polyakov loop, the quark number 
density and the chiral condensate with improved accuracy and $j\to0$ 
extrapolation over earlier publications; the known qualitative features of the low 
temperature phase diagram, which is composed of the hadronic, Bose-Einstein 
condensed (BEC) and BCS phases, are reproduced.  In addition, we newly find 
that around the boundary between the hadronic and BEC phases, nonzero 
quark number density occurs even in the hadronic phase 
in contrast to 
the prediction of the chiral perturbation theory (ChPT), while the diquark 
condensate approaches zero in a manner that is consistent with the 
ChPT prediction.  At the highest $\mu$, which is of order the inverse of
the lattice spacing, all the above observables change drastically, which 
implies a lattice artifact.  Finally, at temperature of order $0.45T_c$, 
where $T_c$ is the chiral transition temperature at zero chemical potential, 
the topological susceptibility is calculated from a gradient-flow method and 
found to be almost constant for all the values of $\mu$ ranging from the hadronic to BCS phase.
This is a contrast to the case of $0.89T_c$ in which the topological susceptibility becomes small as the hadronic phase changes into the quark-gluon plasma  phase.
}

\begin{document} 
\maketitle
\flushbottom

\section{Introduction and summary}
\label{sec:intro}
It is of great importance to probe properties of matter in extreme 
conditions as encountered in neutron stars and relativistic heavy ion
collisions.  In fact, recent observations of the gravitational waves from 
a double neutron star merger event~\cite{TheLIGOScientific:2017qsa}, as well as heavy-ion collision 
experiments performed at the highest energies that can be achieved at 
present~\cite{Andronic:2017pug}, provide empirical information about the equation of state 
(EOS) of matter that occurs in such systems and its evolution.  Since the 
constituents of such matter are strongly coupled, however, one has 
difficulty in accurately evaluating the EOS and transport coefficients 
from QCD.  

Theoretically, all one knows at relatively low temperature 
are the equilibrium properties of matter near zero baryon chemical 
potential and of matter at sufficiently high baryon chemical
potential for weak coupling calculations to work.  At zero baryon 
chemical potential, the chiral condensate and the Polyakov loop 
are known to gradually change with increasing temperature from 
lattice QCD simulations at physical pion mass \cite{Aoki:2009sc,Bazavov:2011nk}. 
The system undergoes a gradual crossover from a hadronic to a 
quark-gluon plasma (QGP) phase around a temperature $T_c$
at which the chiral susceptibility is peaked.  The topological 
susceptibility is also measured on a lattice, which indicates that 
the density of instantons decreases from the value at zero temperature 
to a vanishingly small one \cite{Schafer:1996wv}.   
At nonzero baryon chemical potential, several works based on the multi-parameter reweighting~\cite{Fodor:2001pe,Fodor:2004nz} and the Taylor expansion~\cite{Borsanyi:2012cr} searched for
 a critical end point, but unfortunately it remains to be determined.
Thanks to a sign problem inherent to finite density lattice QCD,
however, the regime in which one can predict from finite density 
QCD is restricted all the way to a regime of such high 
densities as to make perturbative, weak coupling calculations 
reliable.  In this regime, a deconfined superfluid BCS state 
is predicted to occur at sufficiently low temperatures for Cooper pairs 
of diquarks to be formed via one-gluon exchange in the colour antisymmetric 
channel \cite{Alford:2007xm}.

In this work, we examine the temperature and density dependence 
of the phase and topological structure of dense two-colour QCD
by using a lattice simulation with two-flavour Wilson fermions~\footnote{Some very preliminary results have been shown in ref.~\cite{Itou:2018vxf}. }.
In this case, the sign problem can be avoided as follows:  The fundamental 
representation (=quark) of the SU($2$) gauge group takes a (pseudo)real 
representation, so that the determinant of the Dirac operator even at finite 
quark chemical potential $\mu$ is a positive-real or negative-real 
\cite{Muroya:2000qp,Muroya:2002ry,Muroya:2003jp}.  Then, the SU(2) gauge theory coupled to the 
even number of flavours does not suffer from the sign problem.

Even so, 
there is another problem: A numerical instability occurs in the 
low temperature and high density regime.  
This problem comes from a dynamical 
pair-creation and/or pair-annihilation of the lightest hadrons, so that it is theoretically expected 
that the instability appears when $\mu \gtrsim  m_{\rm PS}/2$ is satisfied at 
temperatures close to zero~\cite{Muroya:2000qp,Muroya:2002ry,Muroya:2003jp}.
Here, $\mu$ and $m_{\rm PS}$ denote the quark chemical potential and the pseudoscalar meson mass at zero chemical potential.
One can see how this kind of instability occurs and can be avoided by introducing 
a diquark source in terms of the eigenvalue distribution
of the Dirac matrix \cite{Fukushima:2008su}.
In this work, we thus introduce the diquark source term in the action to 
solve the second problem following refs.~\cite{Kogut:2001na,Kogut:2002cm,Hands:2006ve,Hands:2007uc,Alles:2006ea,Hands:2011hd,Cotter:2012mb, Boz:2013rca,Boz:2015ppa,Braguta:2016cpw} (see section \ref{sec:detail}) and attempt to 
find out the phase structure in the vanishing limit of the source term.

\begin{figure}[h]
\centering\includegraphics[width=10cm]{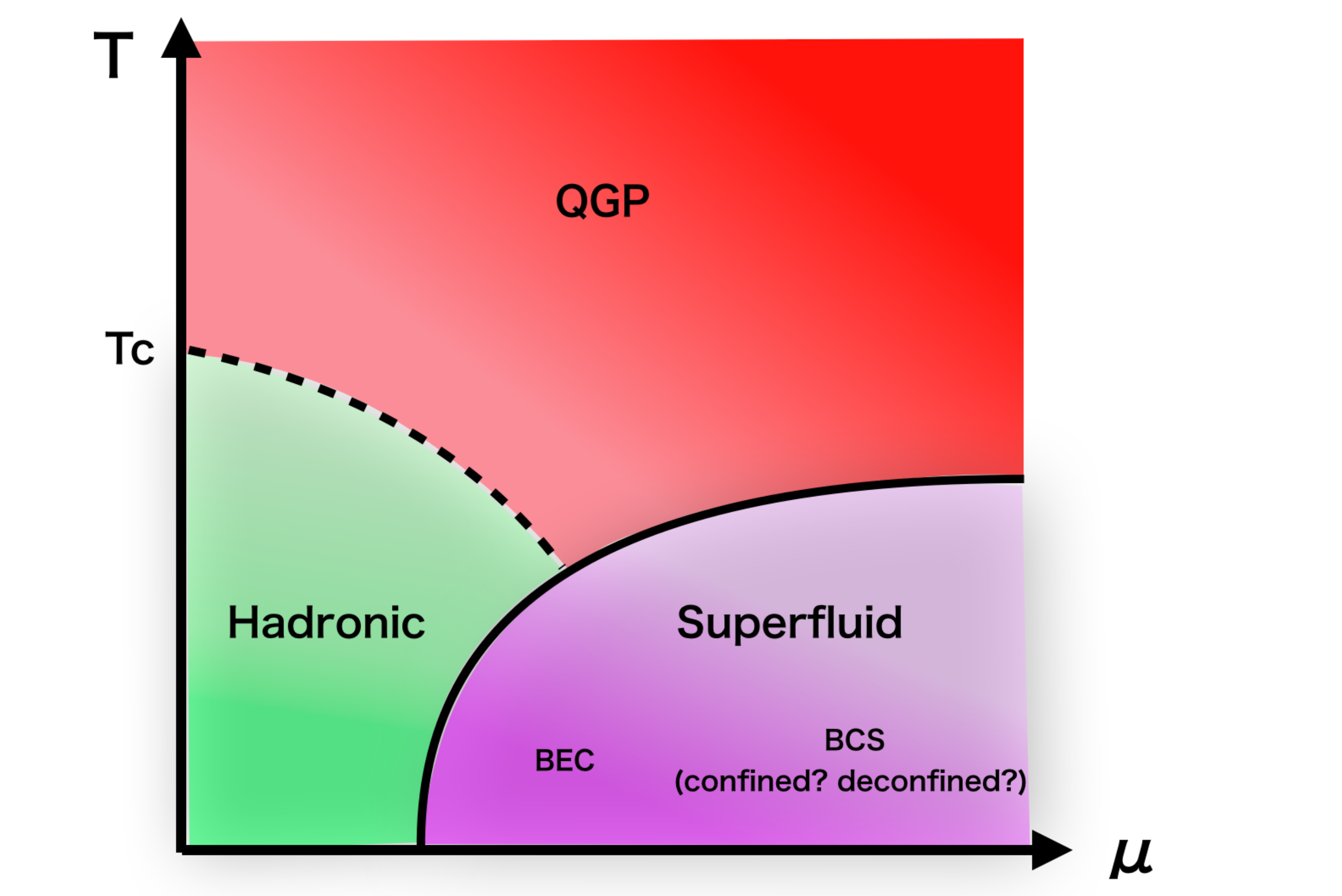}
\caption{Schematic phase diagram of dense two-colour QCD as predicted from earlier works.  
The details of the phases will be given in subsection~\ref{subsec:obs}.}
\label{fig:schematic}
\end{figure}

As for the phase structure of dense two-colour QCD with $N_f=2$, 
earlier lattice simulations~\cite{Hands:2006ve,Hands:2011hd,Cotter:2012mb,Hands:2007uc,Boz:2013rca,Boz:2015ppa,Braguta:2016cpw}
including the diquark source parameter $j$
in the action provided low temperature data for the diquark condensate, 
the Polyakov loop, the quark number density and the chiral condensate 
as a function of quark chemical potential $\mu$.
By taking the limit of $j\to0$, a hadronic phase with vanishing diquark 
condensate was found to change into a superfluid phase with nonzero
diquark condensate around $\mu\sim m_{\rm PS}/2$. 
The phase diagram predicted from the earlier analyses can be 
schematically illustrated in figure \ref{fig:schematic}.
%%%%%%%%%%%%%%%%%%
%%%%%%%%%%%%%%%%%%
We remark in passing that
the two-colour system has baryons as bosons that are degenerate with
mesons in vacuum, a feature clearly different from the usual three-colour 
system~\cite{Kogut:1999iv,Kogut:2000ek,Splittorff:2000mm}.

Consistency of the $\mu$ and $j$ dependence of various quantities 
with the mean-field prediction by the chiral perturbation theory (ChPT)~\cite{Kogut:1999iv,Kogut:2000ek,Kogut:2001na,Adhikari:2018kzh} was examined in detail.  Since the ChPT is applicable 
at sufficiently small $\mu$, it can at most predict the appearance of
a diquark Bose-Einstein condensed (BEC) phase around 
$\mu\sim m_{\rm PS}/2$ and the $\mu$ dependence of the diquark and chiral
condensates.  At still higher densities, it was found that the BCS phase, 
which is characterized by the presence of the Fermi sphere in the 
underlying normal phase, takes over from the diquark BEC phase.  
However, it is still 
controversial whether or not one could see deconfinement at sufficiently 
low temperature.

It is well-known that the two-colour system at the finite density has a similar property to three-colour QCD with isospin chemical potential~\cite{Rischke:2000cn}.
In the three-colour QCD, it is expected that the first order confinement-deconfinement transition occurs at nonzero temperature even at asymptotically 
high densities~\cite{Rischke:2000cn,Son:2000by,Sannino:2002re}.
The intuitive picture of the first order phase transition is as follows:
In the intermediate temperature and high density regime, a typical momentum of quarks is higher than the size of the Fermi surface ($\mu$) or zero-temperature diquark gap ($\Delta_0$) and hence the quarks behave like free particles, while at extremely low temperature the typical momentum of quarks is smaller than $\mu$ or $\Delta_0$ and hence the quarks are quenched. 
As is well-known, the quenched QCD at low temperature exhibits the confinement, so that the low temperature and high density regime must be described by the confining theory.    
By analogy, there could be a similar phase transition by changing temperature in a high density regime~\cite{Schafer:2002yy}.
Since two-colour QCD at finite temperature exhibits the second order confinement-deconfinement phase transition, the transition might be weaker than the three-colour case.

An important and nonperturbative quantity related to the phase structure is the topology.
At low temperatures and intermediate densities, theoretical works are 
based mostly on low energy effective theories \cite{Fukushima:2010bq}.  
Some of the low energy effective theories focus on instantons since such topological 
objects are closely related to the vacuum structure.  In fact, the 
instanton-induced interaction among quarks can promote the chiral phase 
transition as well as the superfluid phase transition \cite{Rapp:1997zu}, 
just like the one-gluon exchange interactions at extremely high densities.
The key question here is how the density of instantons depends on the 
baryon chemical potential.  To answer this question, however, 
nonperturbative calculations are indispensable.  According to the BCS 
theory, the zero-temperature diquark gap ($\Delta_0$) is related to the critical 
temperature ($T_c^{\rm SF}$) for the transition from the superfluid to QGP phase at given chemical potential as
\beq
\Delta_0 = \frac{\pi T_c^{\rm SF}  }{e^{\gamma_E}},\label{eq:BCS-relation}
\eeq
where $\gamma_E$ denotes Euler's constant~\cite{Rapp:1997zu}.
If there are some topological objects in the superfluid phase and they 
increase the diquark gap at zero temperature~\footnote{It is notable that the diquark gap from the Dirac operator has been also discussed in refs.~\cite{Kanazawa:2011tt,Kanazawa:2012zr}.}, therefore, the phase diagram 
is expected to be changed from the perturbative prediction at intermediate 
densities.

As for lattice approach to the topological properties of dense two-colour 
QCD, the earliest work was done for $N_f=4$ and $8$ 
\cite{Alles:2006ea,Hands:2011hd}, 
which shows that the topological susceptibility decreases with $\mu$
in the high density regime, while the result of $N_f=2$ is available only 
on a rather coarse lattice \cite{Hands:2011hd}, 
which suggests that it might be almost constant in the high density regime~\footnote{In the case of three-colour QCD with isospin chemical potential,  $\mu$-independece of the eigenvalue for the overlap Dirac operator has been reported in ref.~\cite{Bali:2016nqn}.}.  
However, the temperature and phase
dependences of the topological susceptibility were not considered.  Since it 
is well-known that the property of the topological distribution depends 
strongly on the temperature $T$ at $\mu=0$ \cite{Schafer:1996wv}, it is 
important to study the topology at 
different temperatures with $N_f$ fixed.  
In this work, therefore, we will investigate the phase 
structure at two relatively low temperatures, $T=0.45T_c$ (section~\ref{sec:lowTphase}) and $T=0.89T_c$ (section~\ref{sec:highTphase}), 
where $T_c$ denotes the chiral transition temperature at $\mu=0$.  
After that, we will study the density dependence of the 
topological susceptibility (see section~\ref{sec:topo}).

%%%%%%%%%%%%%%%%%%
\begin{figure}[h]
\centering\includegraphics[width=10cm]{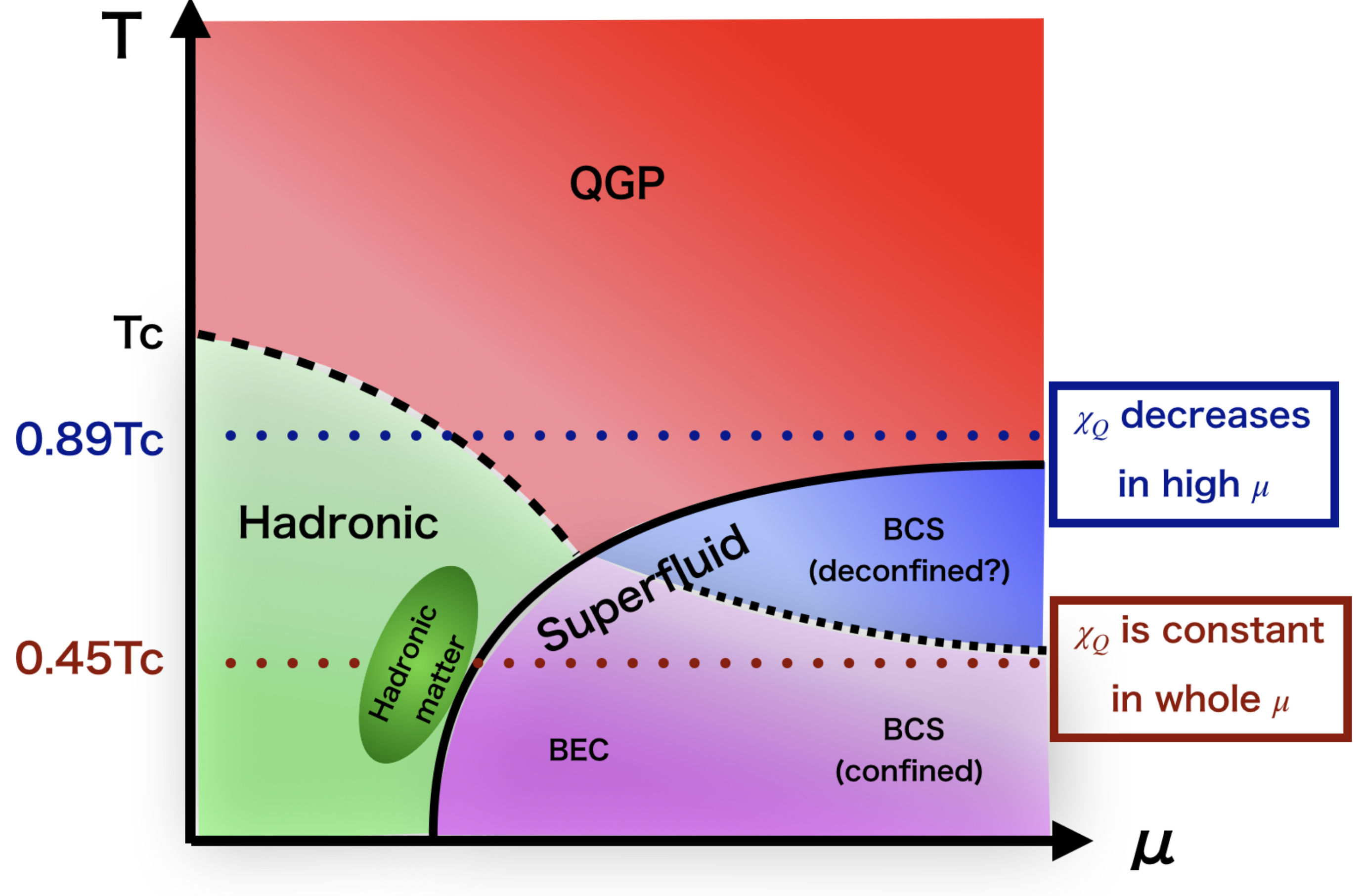}
\caption{Summary of the phase diagram and the topological susceptibility ($\chi_Q$) of dense two-colour QCD revealed by this work.
The BCS (deconfined) phase has yet to be found in this work, and will be investigated in future work.}
\label{fig:summary}
\end{figure}
%%%%%%%%%%%%%%%%%%
We summarize the main conclusions of this paper, which are schematically displayed in figure~\ref{fig:summary}.  By systematically 
measuring the diquark condensate, the Polyakov loop, the quark number density
and the chiral condensate with improved accuracy and $j\to0$ extrapolation 
over earlier investigations~\cite{Hands:2006ve,Cotter:2012mb,Braguta:2016cpw}, 
the known qualitative features of the low temperature phase diagram are 
reproduced at $T=0.45T_c$.  
In improving the $j\to0$ extrapolation, we have developed a new reweighting 
method.  In addition, we newly find that around the boundary between the 
hadronic and BEC phases, so-called the hadronic matter 
regime\footnote{We dare to refer to a hadronic phase with nonzero 
quark number density as ``hadronic matter" because this phase is reminiscent 
of a nuclear medium in real QCD.}, 
nonzero quark number density occurs 
even in the hadronic phase in contrast to the ChPT prediction, while the 
diquark condensate approaches zero in a manner that is consistent with the 
ChPT prediction.  At the highest $\mu$, which is of order the inverse of
the lattice spacing, all the above observables suffer from a strong lattice artifact and change drastically.
  There is no clear evidence for 
deconfinement at densities below the regime where a lattice artifact 
takes effect; this is consistent with the analysis with staggered fermions 
\cite{Braguta:2016cpw}, but at odds with that with Wilson fermions 
\cite{Cotter:2012mb}.  Finally, 
at temperature of $0.45T_c$, 
the topological susceptibility is calculated from a gradient-flow method 
\cite{Luscher:2010iy} and 
found to be almost constant for all the values of $\mu$ ranging from the hadronic to BCS (confined) phase. 
This is a contrast to the case of $0.89T_c$ in which the topological 
susceptibility becomes small as the hadronic phase changes into the 
QGP phase.

\section{Simulation detail}
\label{sec:detail}
In this section we start with the lattice action.  We then define the 
physical quantities to be measured and show how to determine the phases 
from the behaviour of the quantities.  We finally set the simulation 
parameters.

\subsection{Lattice action}
\label{subsec:action}
In this work, as a lattice gauge action, we utilize the Iwasaki gauge 
action, which is composed of the plaquette term with $W^{1\times 1}_{\mu\nu}$ 
and the rectangular term with $W^{1\times 2}_{\mu\nu}$,  
\beq
S_g = \beta \sum_x \left(
 c_0 \sum^{4}_{\substack{\mu<\nu \\ \mu,\nu=1}} W^{1\times 1}_{\mu\nu}(x) +
 c_1 \sum^{4}_{\substack{\mu\neq\nu \\ \mu,\nu=1}} W^{1\times 2}_{\mu\nu}(x) \right) ,
\eeq
where $\beta=4/g_0^2$ in the two-colour theory, and $g_0$ denotes the bare 
gauge coupling constant.  The coefficients $c_0$ and $c_1$ are set to 
$c_1=-0.331$ and $c_0=1-8c_1$.

As a lattice fermion action, we use the two-flavour Wilson fermion 
action including the quark number operator and the diquark source term, which 
is given by
\beq
S_F= \bar{\psi}_1 \Delta(\mu)\psi_1 + \bar{\psi}_2 \Delta(\mu) \psi_2 - J \bar{\psi}_1 (C \gamma_5) \tau_2 \bar{\psi}_2^{T} + \bar{J} \psi_2^T (C \gamma_5) \tau_2 \psi_1.\label{eq:action}
\eeq
Here, the indices $1,2$ denote the flavour label, and $\mu$ is the 
quark chemical potential.  The additional parameters $J$ and $\bar{J}$ 
correspond to the anti-diquark and diquark source parameters, respectively.
For simplicity, we put $J=\bar{J}$ and assume that it takes a real value. 
Note that $J=j \kappa$, where $j$ is a source parameter in the corresponding 
continuum theory.  The factor $\kappa$ comes from the rescaling of the Wilson 
fermion on the lattice as will be specified below.  The $C$ in the last 
two terms is the charge conjugation operator, and $\tau_2$ acts on the 
colour index.  Finally, $\Delta(\mu)$ is the Wilson-Dirac operator 
including the number operator, which is explicitly given by
\beq 
\Delta(\mu)_{x,y} = \delta_{x,y} 
&&- \kappa \sum_{i=1}^3  \left[ ( 1 - \gamma_i)  U_{x,i}\delta_{x+\hat{i},y} + (1+\gamma_i)  U^\dagger_{y,i}\delta_{x-\hat{i},y}  \right]\\
&&- \kappa   \left[ e^{+\mu}( 1 - \gamma_4)  U_{x,4}\delta_{x+\hat{4},y} + e^{-\mu}(1+\gamma_4)  U^\dagger_{y,4}\delta_{x-\hat{4},y}  \right],
 \eeq 
where $\kappa$ is the hopping parameter.

Now, the action (\ref{eq:action}) contains three types of fermion 
bilinears: $\bar{\psi} \psi$, $\bar{\psi} \bar{\psi}$ and $\psi \psi$.  
To build a single kernel matrix from the fermion action, 
we introduce an extended fermion matrix ($\mathcal M$) as
\beq
S_F&=& (\bar{\psi}_1 ~~ \bar{\varphi}) \left( 
\begin{array}{cc}
\Delta(\mu) & J \gamma_5 \\
-J \gamma_5 & \Delta(-\mu) 
\end{array}
\right)
\left( 
\begin{array}{c}
\psi_1  \\
\varphi  
\end{array}
\right)
 \equiv  \bar{\Psi} {\mathcal M} \Psi,  \label{eq:def-M}
\eeq
where
$\bar{\varphi}=-\psi_2^T C \tau_2, ~~~ \varphi=C^{-1} \tau_2 \bar{\psi}_2^T.$
The square of the extended matrix can be diagonal because 
$J(=\bar{J})$ is real.   We thus obtain
\beq
\det[{\mathcal M}^\dag {\mathcal M}] = 
\det[ \Delta^\dag(\mu)\Delta(\mu) + J^2 ] \det[  \Delta^\dag (-\mu) \Delta(-\mu) + J^2  ]. \label{eq:MdagM}
\eeq
Note that $\det[{\mathcal M}^\dag {\mathcal M}]$ corresponds to the fermion 
action for the {\it four-flavour} theory, since a single $\mathcal{M}$ in 
eq.\ (\ref{eq:def-M}) represents the fermion kernel of the two-flavour 
theory.  To reduce the number of fermions, we take the square root of 
the extended matrix in the action and utilize the Rational Hybrid Montecarlo 
(RHMC) algorithm in our numerical simulations.

\subsection{Observables and definition of phases}
\label{subsec:obs}
Now, we turn to the notation of the phases as described in figure 
\ref{fig:schematic}.  Following earlier investigations 
\cite{Braguta:2016cpw}, 
we use the name of each phase as shown in table \ref{table:phase}.
%%%%%%%%%%%%%%
\begin{table}[h]
\begin{center}
\begin{tabular}{|c||c|c|c|c|c|}
\hline
 \multicolumn{1}{|c||}{}  & \multicolumn{2}{c|}{Hadronic} & \multicolumn{1}{c|}{QGP}  &  \multicolumn{2}{c|}{Superfluid}  \\  
\cline{3-3} \cline{5-6}  & & Hadronic matter & & BEC & BCS \\  
 \hline \hline
$\langle |L| \rangle$ & zero  & zero  & non-zero &   &   \\
$\langle qq \rangle$ & zero  &  zero  & zero & non-zero & $\propto \mu^2$ \\ 
$ \langle n_q \rangle $ &  $ \langle n_q \rangle = 0$ &  $ \langle n_q \rangle  > 0$  & $\langle n_q \rangle \ge 0$ & non-zero & $\langle n_q \rangle /n_q^{\mbox{tree}} \approx 1$ \\ 
 \hline
\end{tabular}
\caption{ Definition of the phases } \label{table:phase}
\end{center}
\end{table}
%%%%%%%%%%%%%
Here, $L$ is the magnitude of the Polyakov loop, 
\beq
L= \frac{1}{N_s^3} \sum_{\vec{x}} \prod_\tau U_{4} (\vec{x}, \tau),
\eeq
which plays the role of an approximate order parameter for confinement.
$\langle qq \rangle$ denotes the diquark condensate, 
\beq
\langle qq \rangle \equiv \frac{\kappa}{2} \langle \bar{\psi}_1 K \bar{\psi}_2^T - \psi_1 K \psi_2^T \rangle,
\eeq 
where $K=C \gamma_5 \tau_2$.  This
is nothing but the order parameter for superfluidity.  
$n_q$ denotes the quark number density,
\beq
a^3 n_q= \sum_{i} \kappa \langle \bar{\psi}_i (x) (\gamma_0 -\mathbb{I}_4) e^\mu U_{4} (x) \psi_i (x+\hat{4})  + \bar{\psi}_i (x) (\gamma_0 + \mathbb{I}_4) e^{-\mu}U_4^\dag (x-\hat{4} )\psi_i (x-\hat{4})) \rangle,\nonumber\\
\eeq
which is the time-like component of a conserved current.
Note that this quantity does not require a renormalization and 
goes to $2N_f N_c$ in the high $\mu$ limit on the lattice.
Incidentally, $n_q^{\mbox{tree}}$ is the quark number density 
defined by the free field on a finite lattice \cite{Hands:2006ve}:
\beq
n_q^{\mbox{tree}}(\mu) = \frac{4N_cN_f}{N_s^3 N_\tau} \sum_k \frac{i \sin \tilde{k}_0 [ \sum_i \cos k_i -\frac{1}{2\kappa} ]}{[\frac{1}{2\kappa} -\sum_\nu \cos \tilde{k}_\nu ]^2 +\sum_\nu \sin^2 \tilde{k}_\nu},
\eeq
where
\beq
\tilde{k}_0 = k_0 -i\mu = \frac{2\pi}{N_\tau} (n_0+1/2) -i\mu,~~~~~~~~\tilde{k}_i = k_i = \frac{2\pi}{N_s}n_i,~~~~i=1,2,3 .
\eeq
In the continuum limit, $n_q^{\mbox{tree,cont}}=\mu^3/(3\pi^2)$.

Note that we use the definition in table \ref{table:phase}
no matter whether the system is at zero temperature or not.  In fact, some 
``non-zero" quantities could come from thermal effects.  For instance, the 
finiteness of $n_q$ in the ``hadronic matter'' phase 
is expected to be one of them as will be discussed later.  Although 
this hadronic matter is just a part of the ordinary hadronic phase,
we use this terminology to distinguish between the hadronic state with
$n_q>0$ and the hadronic vacuum.

The nonzero value of the diquark condensate indicates that the system is 
in a superfluid state.  We expect that a low density regime in the 
superfluid phase could be described  in terms of a diquark BEC picture.
In this picture, the diquark ($q$--$q$) pairs behave as bosons that form
a condensate when a typical correlation length of the system, namely, the 
coherence length of the pair, is sufficiently shorter than the  
inter-quark spacing $\sim\mu^{-1}$.  This is the case with a rarefied system.
With increasing density, however, the inter-quark spacing becomes
shorter and shorter.  If the spacing is much shorter than the typical 
correlation length, then, the number of the pairs significantly decreases
since only quarks near the Fermi surface that would occur in the absence of 
attraction between quarks can undergo Cooper pairing.  At the same time, the 
Cooper pairs spatially overlap with each other, which gives rise to a coherent
state.  It is nevertheless noteworthy that most of the quarks behave like 
free particles.  For this phase, generally referred to as the BCS 
phase, the quark number density can thus be estimated from 
the free particle propagator including the chemical potential $\mu$.  We thus 
expect that even on a lattice, the quark number density can be 
approximated by $n_q^{\mbox{tree}}$ in the BCS phase.  Furthermore, we expect 
that $\langle qq \rangle $ is proportional to $\mu^2$, which corresponds 
to the area of the Fermi surface.

It might be also tempting to consider the chiral condensate, $\langle \bar{q} q \rangle \equiv \langle \bar{\psi}_i \psi_i \rangle$ for one flavour, as an order 
parameter of chiral symmetry breaking.  In our numerical calculations, 
however,  we utilize the massive Wilson fermions, which explicitly break the 
chirality.  Although we will discuss numerical results for the chiral 
condensate in each phase, we will not use the quantity for the definition 
of the phase.

We conclude this subsection by mentioning that a lattice artifact
can induce a peculiar phase in the high density regime.  There are two 
possible origins that can result in such a phase in the numerical 
simulations.  The first one can be found by observing the behaviour of the
diquark condensate.  In fact, when the value of $\mu$ approaches $1$ 
in lattice unit, the propagation of the quarks is quenched, 
leading to a significant decrease in the amplitude of $\langle qq \rangle$ 
with increasing $\mu$.  This kind of behaviour was 
reported in refs.\ \cite{Kogut:2002cm,Braguta:2016cpw} in the $N_f=2,4$
simulations using the staggered fermions.

The second origin is the finite volume effect,  which manifests itself
when all the lattice site are occupied by the quarks.
The signal of a peculiar phase coming from the finite volume effect 
can be found by observing the saturation of $n_q$ in lattice unit, 
which corresponds to $a^3 n_q=8$ in our notation.
In the actual simulation of this work, the maximal value of $ a^3 \langle n_q \rangle $ is 0.707(2), 
which is significantly smaller than $8$.  We can thus safely avoid 
such an artifact.

\subsection{Simulation parameters}
\label{subsec:para}
We perform the simulations with ($\beta, \kappa, N_s, N_\tau$) $=$ 
($0.800,0.159,16,16$) and ($0.800,0.159,32,8$).
According to preliminary results in ref.~\cite{Lee}, 
($\beta,\kappa$) $=$ ($0.800, 0.159$) gives  $m_{\rm PS}/m_V=0.823(9)$ and 
$am_{\rm PS}= 0.623(3)$, where $m_V$ is the vector meson mass in vacuum.  
Furthermore, the chiral phase transition 
at $\mu=0$ occurs at $(\beta, N_\tau$) $=$ ($0.900,10$) with the simulations 
performed at fixed $m_{\rm PS}/m_{V}$. Eventually, the scale setting 
utilizing $w_0$~\cite{Borsanyi:2012zs} in the gradient flow method~\cite{Luscher:2010iy} leads to $T=0.45T_c$ and 
$0.89T_c$, respectively, for the former and latter simulations.

\section{Simulation results: Phase diagram at $T=0.45T_c$}
\label{sec:lowTphase}
Let us proceed to show the finite density phase diagram at fixed 
temperature ($T=0.45T_c$) by analysing the $\mu$ dependence of
the observables described in subsection \ref{subsec:obs}.

\subsection{Polyakov loop}
\label{subsec:pol}
%%%%%%%%%%%%%%%%%%
\begin{figure}[h]
\centering\includegraphics[width=15cm]{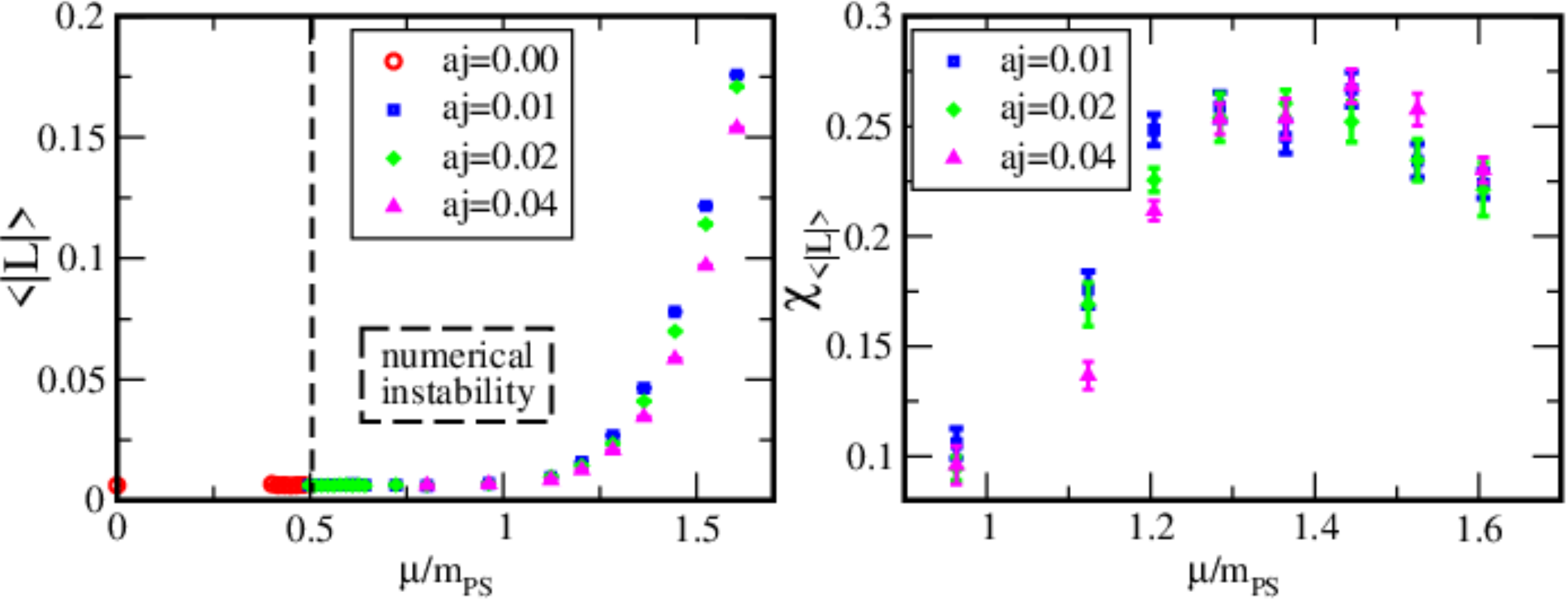}
\caption{(Left) The $\mu$ and $j$ dependence of the Polyakov loop obtained for
$\beta=0.800, 16^4$ lattices. (Right) The susceptibility of the Polyakov loop 
in a high $\mu$ regime. }
\label{fig:Ploop}
\end{figure}
%%%%%%%%%%%%%%%%%%
We start with the Polyakov loop, whose $\mu$-dependence is depicted in 
figure \ref{fig:Ploop}.  The symbols (colours) denote the different values of the diquark
source ($j=J/\kappa$) in lattice unit; 
the data with circle (red), square (blue), diamond (green) and triangle (magenta) are 
generated by the $aj=0.00, 0.01, 0.02$ and $0.04$ simulations, respectively.
Note that the theoretical threshold $\mu$ value of the numerical instability
is $\mu/m_{\rm PS}=1/2$, beyond which the lightest hadrons are 
frequently created and annihilated.  In fact, we can implement the HMC 
simulation without the diquark source for $\mu/m_{\rm PS} < 0.50$, while the 
Metropolis test in the HMC simulation is not accepted even within a tiny 
Montecarlo step ($\sim 1/1000$) for $\mu/m_{\rm PS} \ge 0.50$.  In the 
latter regime, we thus have to utilize the RHMC algorithm including the 
diquark source in the action.

As can be seen from the left panel of figure \ref{fig:Ploop}, the Polyakov 
loop takes on a nonzero value for $\mu/m_{\rm PS} \gtrsim 1.2$.
Normally, it is a signal of the transition or crossover from the 
confined to deconfined phase.
The right panel of figure \ref{fig:Ploop} shows the susceptibility of the 
Polyakov loop obtained for $aj=0.01$--$0.04$.  The susceptibility is marginally 
peaked at $\mu=\mu_D$ with $\mu_D /m_{\rm PS}\sim 1.44$.
A clear $j$-dependence of the peak position is not observed in our calculation.
Since $a\mu_D$ is close to unity ($a\mu_D = 0.90$), the obtained structure of the Polyakov 
loop might be associated with a lattice artifact as we shall see.

\subsection{Diquark condensate}
\label{subsec:diquark}
The other quantity that helps determine the phase diagram is the 
expectation value of the diquark condensate, which have been measured by utilizing the noise method for each configuration.  
It is however 
difficult to evaluate this expectation value from the extrapolation of 
$j\rightarrow 0$ limit, as is the case with earlier investigations
\cite{Hands:2006ve,Cotter:2012mb,Braguta:2016cpw}.
Let us then propose a reweighting method with respect
to $j$ at fixed $\beta,\mu,\kappa$ ~\footnote{A similar reweighting method 
was discussed in the context of QCD with nonzero isospin chemical 
potential \cite{Brandt:2017oyy}.}.  In this method, the reweighting 
factor from the original lattice parameters ($\beta_0,\kappa_0,\mu_0,j_0$) 
to the measured parameters ($\beta,\kappa,\mu,j$) is given by
\beq
R_j &\equiv&  \frac{\det [D (\kappa,\mu,j)]}{\det [D (\kappa_0,\mu_0,j_0)]}  e^{-(\beta-\beta_0) S_g[U]} .
\eeq
In our simulation, the value of $j$ alone is changed between the 
configuration generation and the measurement of the observables.
Thus, the reweighting factor explicitly reads
\beq
R_j &=& \left(  \frac{\det [\Delta^\dag (\mu) \Delta(\mu)+ J^2)]  \det [\Delta^\dag (-\mu) \Delta(-\mu)+ J^2)] }{\det [\Delta^\dag (\mu) \Delta(\mu)+ J_0^2)] \det [\Delta^\dag (- \mu) \Delta(-\mu)+ J_0^2)]} \right)^{1/2} \nonumber\\
&=& \left( \det [1+ (J^2-J_0^2)(\Delta^\dag (\mu) \Delta(\mu)+J_0^2)^{-1})]  \det [1+ (J^2-J_0^2) (\Delta^\dag (-\mu) \Delta(-\mu)+J_0^2)^{-1})] \right)^{1/2}. \nonumber\\
\eeq
In our calculations, the value of $J^2=(j\cdot \kappa)^2$ is typically
$O(10^{-6})$ in lattice unit, so that the Taylor expansion,
 $\det [1+A] = e^{\mbox{Tr}\ln [1+A] } \sim 1+\mbox{Tr} A$, is valid.
In fact, for $\mu/m_{\rm PS} < 0.50$ ($a\mu \le 0.30$), the data for $aj  > 0$ are obtained 
by the reweighting from the configurations generated with $aj_0=0.00$.
On the other hand, for $\mu/m_{\rm PS} \ge 0.50$ ($a\mu \ge 0.31$), the data for $aj \le 0.01$ are measured by using the reweighting method
from the configurations with $aj_0=0.01$.
 It is numerically confirmed that the leading correction of the reweighting 
factor ($R_j-1$) is less than $10^{-3}$ in our calculations.

%%%%%%%%%%%%%%%%%%
\begin{figure}[h]
\vspace{0.5cm}
\centering\includegraphics[width=15cm]{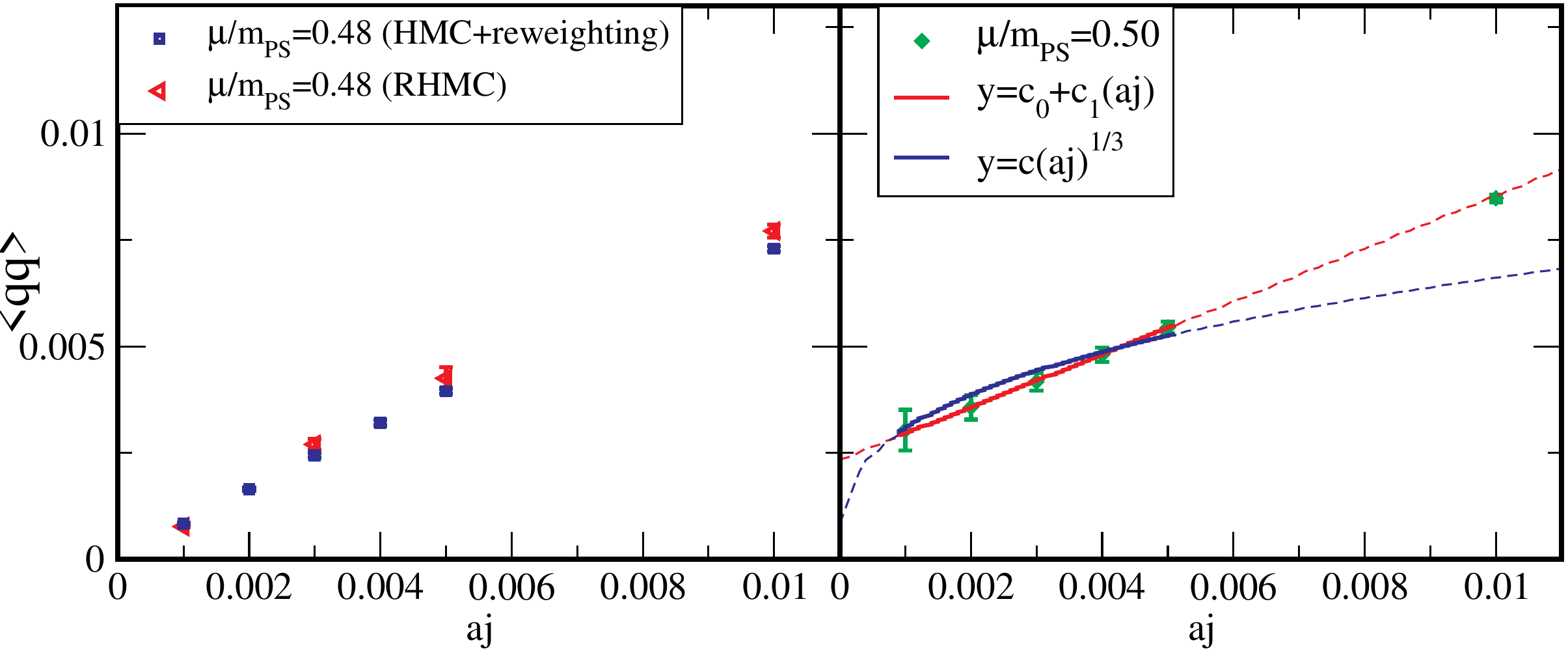}
\caption{(Left) The $j$-dependence of the diquark condensate for $\mu/m_{\rm PS}=0.48$, which is around the boundary of changing the algorithm and measurement methods. The square (blue) data generated by the HMC algorithm and reweighting method from $aj=0.00$, while triangle (red) data obtained by the RHMC algorithm with each $j$ term in the action.
(Right) The $j$-dependence of the diquark condensate for $\mu/m_{\rm PS}=0.50$, which is fitted to two different
functions.  The diamond (green) data generated by the RHMC algorithm and 
reweighting method from $aj=0.01$.  Here the fitting rage is limited to
$0.001 \le aj \le 0.005$ in (beyond) which the fitting is drawn 
in solid (broken) line.}
\label{fig:diquark-0.30}
%\vspace{6cm}
\end{figure}
%%%%%%%%%%%%%%%%%%
We firstly investigate the consistency of the data obtained by the different methods around $\mu/m_{\rm PS} = 0.50$. 
In the left panel of figure~\ref{fig:diquark-0.30}, the square (blue) data for $\mu/m_{\rm PS}=0.48$ are generated by the HMC algorithm without the $j$ term in the action and are measured by using the reweighting method from the configurations with $aj=0.00$, while the triangle (red) data for the same $\mu/m_{\rm PS}=0.48$ are obtained by the configurations generated by using the RHMC algorithm for each value of $j$ and are also measured using the same value of $j$. 
Both data for each $j$ are $3\sigma$ consistent, and furthermore both extrapolated values of $\langle qq \rangle $ in the $j \rightarrow 0$ limit vanish.
We also show the data for $\mu/m_{\rm PS}=0.50$ ($a\mu=0.31$) in the right panel, where the data are generated by the RHMC algorithm with $aj=0.01$, since as explained, at this $a\mu$ it is very hard to generate the configurations utilizing the HMC without the $j$ term.
The diquark condensate for $\mu/m_{\rm PS} \ge 0.50$ are measured by the reweighting method from $aj=0.01$.
At this critical $\mu$, the ChPT predicts that
the diquark condensate behaves like $j^{1/3}$, and  
in the case of the
study with staggered fermions, the numerical evidence was obtained~\cite{Braguta:2016cpw}.  In this work, however, the linear extrapolation in $j$ 
rather than the $j^{1/3}$ scaling can 
reproduce the data beyond the fitting range of $aj$ with a reasonable 
chi-square, and hence there seems no clear evidence for the $j^{1/3}$ scaling
in our simulation.

%%%%%%%%%%%%%%%%%%
\begin{figure}[h]
\vspace{0.5cm}
\centering\includegraphics[width=15cm]{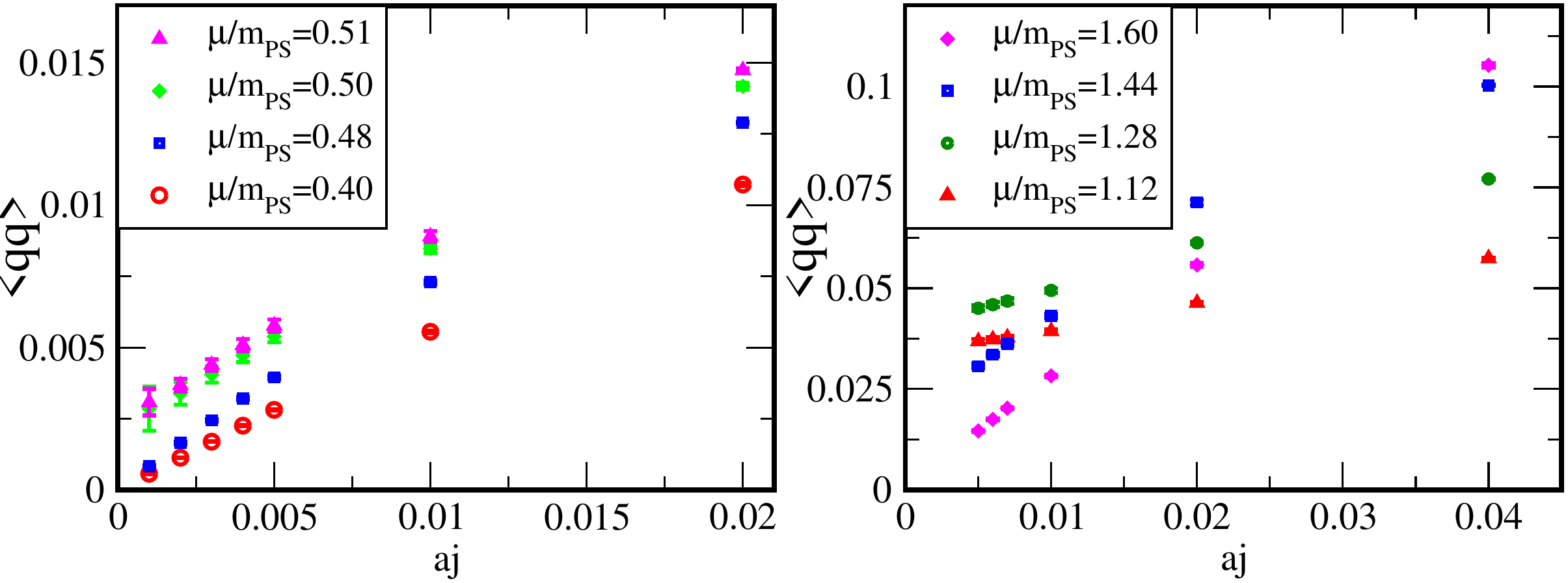}
\caption{The $j$-dependence of the diquark condensate for several 
$\mu/m_{\rm PS}$.}
\label{fig:diquark}
%\vspace{6cm}
\end{figure}
%%%%%%%%%%%%%%%%%%
The left and right panels of figure \ref{fig:diquark} present how the diquark 
condensate depends on $j$ in the low $\mu$ and high $\mu$ regimes, 
respectively.  
In the $j\to0$ limit, the left panel shows that the data 
go to zero for $\mu/m_{\rm PS} < 0.50$, as denoted by circle (red) and 
square (blue) symbols, while having a nonzero expectation value for 
$\mu/m_{\rm PS} \ge 0.50$, as denoted by diamond (green) and triangle (magenta) 
symbols.  The critical value, $\mu/m_{\rm PS}=0.50$, corresponds to the 
onset of the numerical instability,  at which superfluidity is known to
occur \cite{Fukushima:2008su}.

In the high $\mu$ regime, as can be seen from the right panel in 
figure \ref{fig:diquark}, the diquark condensate has a larger $j$ dependence
for higher $\mu$.  Normally, the extrapolated value of the diquark 
condensate increases with $\mu$, namely $\langle qq \rangle \propto \mu^2$, in the BCS phase, but it starts to decrease around
$\mu/m_{\rm PS} = 1.28$, which corresponds to $a\mu=0.80$.
This behaviour might be a signal of the strong lattice artifact, as 
discussed in subsection\ \ref{subsec:obs}.
It might be notable that the similar behaviour occurs at high $a\mu$ in refs.~\cite{Kogut:2002cm, Braguta:2016cpw}, though the lattice fermion in both works is the staggered one.

\begin{figure}[h]
\vspace{0.5cm}
\centering\includegraphics[width=15cm]{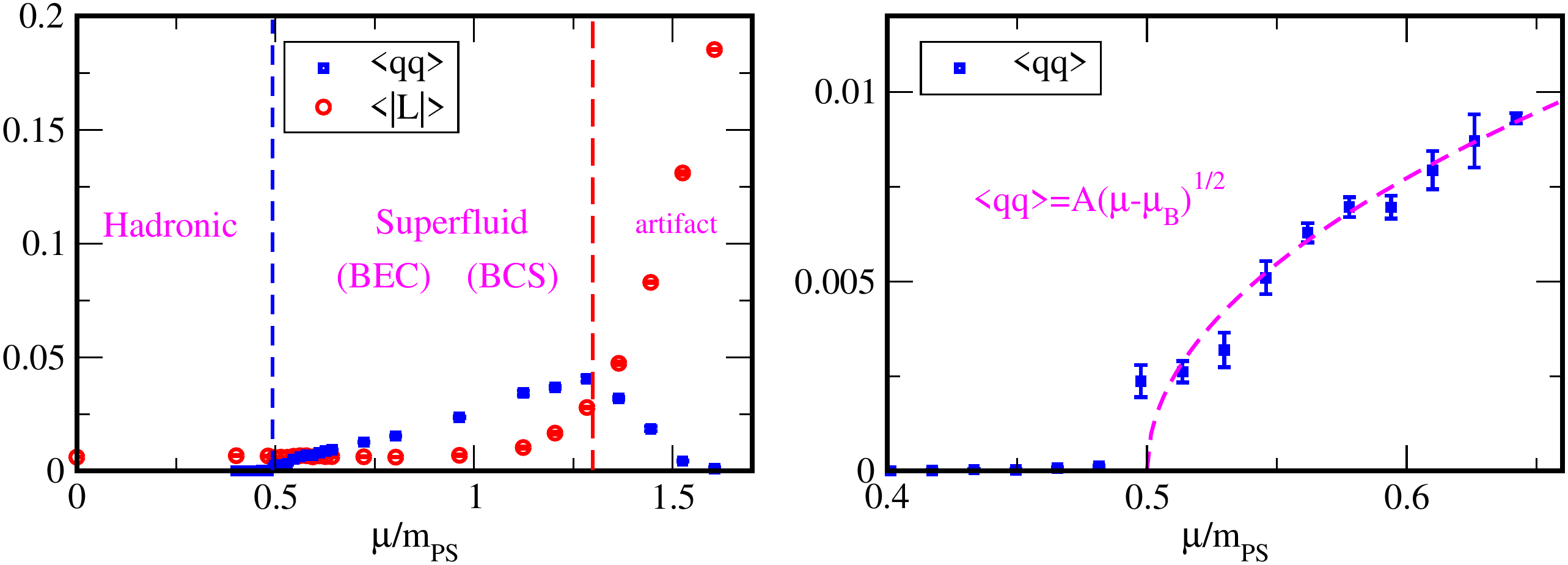}
\caption{(Left) Summary of the data of the Polyakov loop and the diquark 
condensate in the $j\to0$ limit. (Right) The data of the diquark condensate 
in the $j\to0$ limit around the critical point of $\mu=\mu_B=0.50 m_{\rm PS}$.}
\label{fig:phase-diagram}
%\vspace{6cm}
\end{figure}

Now, let us exhibit, in figure \ref{fig:phase-diagram}, the phase 
diagram at $T=0.45T_c$ by using the data of the Polyakov loop and the 
diquark condensate.  
%%%%%%%%%%%%%%%%%%
%%%%%%%%%%%%%%%%%%
According to the definition of the phases given in 
table \ref{table:phase}, the hadronic and  superfluid phases appear
at $T=0.45T_c$.  Although it remains to be determined whether or not the 
superfluid phase is composed of the BEC and BCS phases, it will be addressed
in the next subsection.  Thanks to the 
effectiveness of the reweighting method concerning $j$, we can 
precisely find the location of the transition between the hadronic and superfluid phases as $\mu_B/m_{\rm PS} \simeq 1/2$.  This result is consistent with 
the ChPT prediction that for $\mu \ge m_{\rm PS}/2$ the lightest 
hadrons are frequently created.  
As shown in the right panel in figure~\ref{fig:phase-diagram}, the value of $\langle qq \rangle$ is appreciably non-zero at $\mu/m_{\rm PS}=0.50$. This behaviour may be due to a finite size effect, which is expected to appear when the effective diquark mass is smaller than $1/(a N_s)$.  

The scaling law around the critical point, 
\beq
\langle qq \rangle \propto (\mu -\mu_B)^{\beta_m},
\eeq
 can be investigated. 
Here, we have taken $\mu_B=m_{\rm PS}/2$ and $\beta_m=0.50$, which are the mean-field predictions
by ChPT~\cite{Kogut:2000ek}, since the determination of $\mu_B$ and $\beta_m$ from the numerical data was very hard.  
The chi-square over d.o.f. of the fitting for the $9$ data points in $1/2 \le \mu/m_{\rm PS} <0.70$ is reasonable ($\chi^2/(d.o.f.)=1.31$), and we find that the theoretical curve
based on the scaling law is consistent with the data as shown in the right panel in figure~\ref{fig:phase-diagram}.

\subsection{Quark number density and chiral condensate}
\label{subsec:number}
\begin{figure}[h]
\vspace{0.5cm}
\centering\includegraphics[width=15cm]{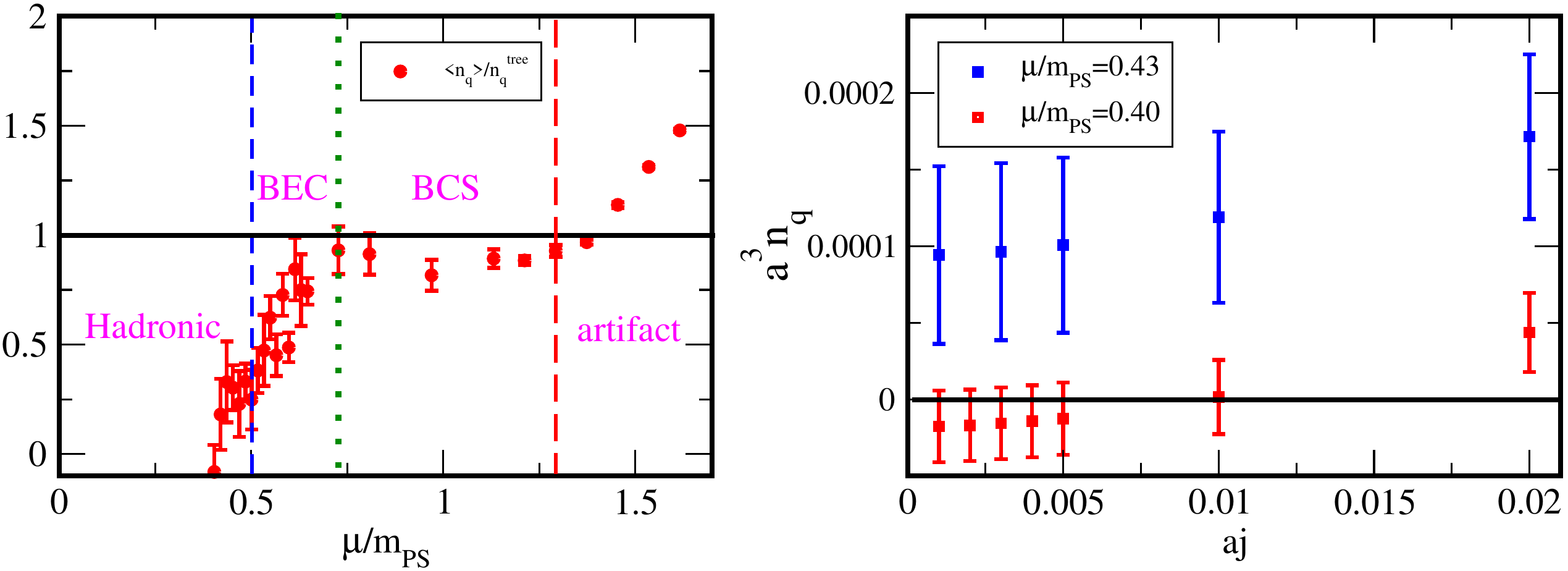}
\caption{ (Left) The quark number density plotted as a function of $\mu$ 
in the $j\to0$ limit. 
 (Right) The $j$ dependence of the quark number density at 
$\mu/m_{\rm PS}=0.40$ and $0.43$.}
\label{fig:number-density}
\end{figure}

We turn to the quark number density and chiral condensate, 
which have been measured by utilizing the noise method. 
The result  for the quark number density is shown in 
figure~\ref{fig:number-density}.
%%%%%%%%%%%%%%%%%%
%%%%%%%%%%%%%%%%%%
We can  observe from the left panel a regime where the lattice data 
for the quark number density $n_q$ are almost consistent with the free 
field behaviour ($n_q^{\mbox{tree}}$).   This regime, which ranges
$0.72 \lesssim \mu/m_{\rm PS} \lesssim 1.28$, may well be identified as 
the weakly coupled BCS phase, rather than the strongly coupled BEC phase, 
which ranges $0.50 \lesssim \mu/m_{\rm PS} \lesssim 0.72$.
Some works based on the Nambu-Jona-Lasinio model have also found a roughly consistent crossover point $\mu/m_{\rm PS} =0.8$--$0.85$ in refs.~\cite{Sun:2007fc,He:2010nb}.

 It is interesting to note that even in the hadronic phase, the
quark number density in the $j\to0$ limit takes on a nonzero value
in a narrow regime just below $\mu/m_{\rm PS}=1/2$,  as can be clearly 
seen from the right panel in figure \ref{fig:number-density}.
In fact, such a regime ranges $0.42 \lesssim \mu/m_{\rm PS} \lesssim 0.50$.
Here, this regime is referred to as ``hadronic matter". 
The presence of the hadronic matter is consistent with a recent Schwinger-Dyson analysis~\cite{Contant:2019lwf}.
This is apparently at odds with the ChPT prediction that $n_q$ becomes 
nonzero when $\mu$ reaches $m_{\rm PS}/2$.  Note, however, that the ChPT
prediction holds for sufficiently low temperature and that the lattice data
have been obtained at $T\simeq 0.1m_{\rm PS}$.  Since the ChPT predicts
that the diquark (antidiquark) mass in the medium is $m_{\rm PS}-2\mu$
$(m_{\rm PS}+2\mu)$, the fact that $T \sim m_{\rm PS}-2\mu$ holds
at $\mu\simeq0.42m_{\rm PS}$ implies that diquarks are thermally 
excited\footnote{Note that finite size effects are expected to 
come in when $1/N_s \simeq (m_{\rm PS}-2\mu)a$, i.e., $\mu\simeq0.45m_{\rm PS}$.
The value of $n_q$ in the immediate vicinity of $\mu=0.50m_{\rm PS}$ may thus
suffer from such effects.}.  
It is thus reasonable for the quark number density to start 
increasing from nearly zero at $\mu\simeq0.42m_{\rm PS}$.  This 
implication could be confirmed by measuring the diquark mass 
in the medium as in refs.\ \cite{Muroya:2002ry, Hands:2007uc,Wilhelm:2019fvp}, 
which will be addressed in the future.

%%%%%%%%%%%%%%%%%%
\begin{figure}[h]
%\vspace{-6cm}
\centering\includegraphics[width=8cm]{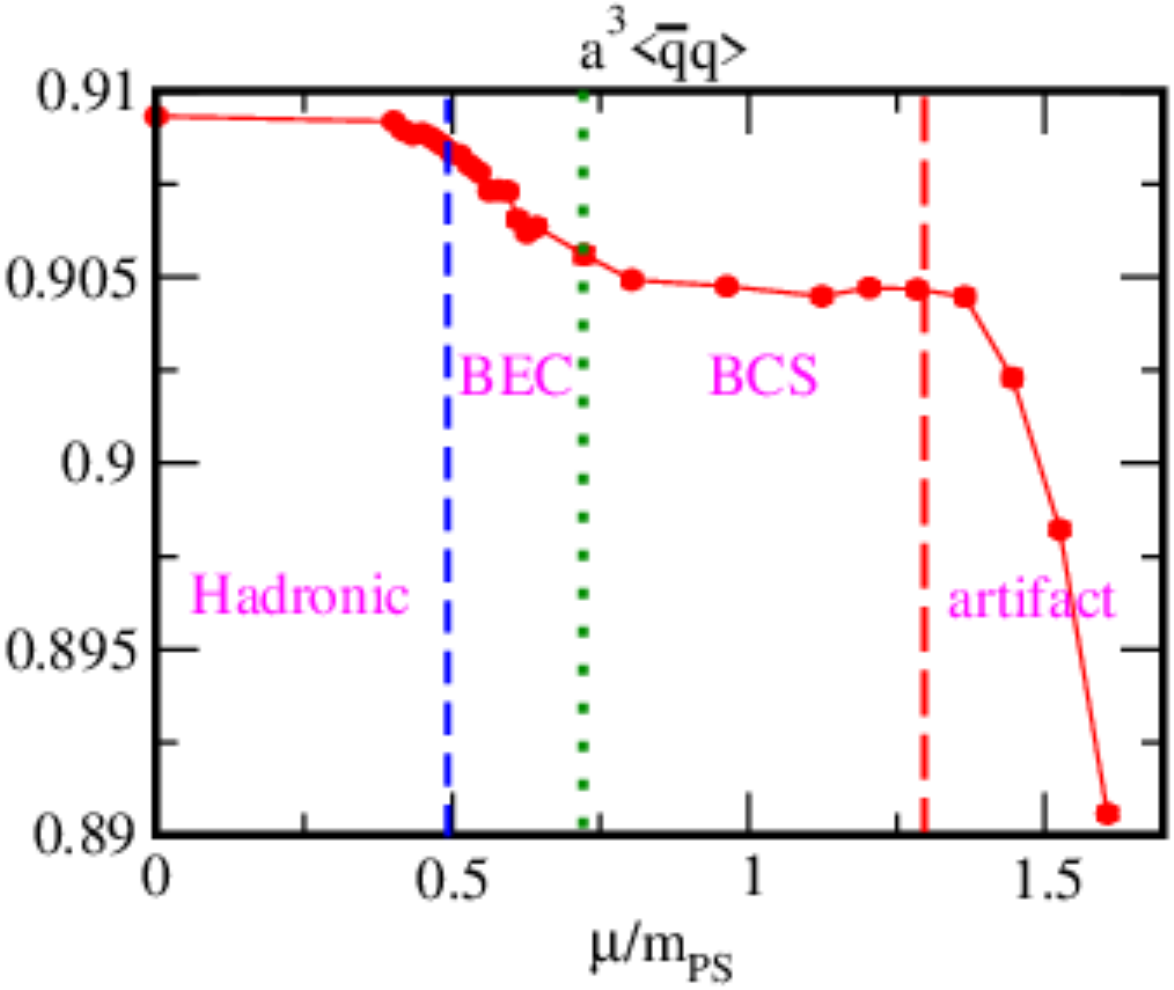}
%\vspace{-1cm}
\caption{The chiral condensate plotted as a function of $\mu$ 
in the $j=0$ limit.  }
\label{fig:chiral-cond}
%\vspace{6cm}
\end{figure}
%%%%%%%%%%%%%%%%%%
Finally, for completeness, we show the chiral condensate in the left 
panel of figure \ref{fig:chiral-cond}.  
Note that here the additive renormalization is not performed.
Interestingly, a plateau appears
in the BCS phase ($0.72 \lesssim \mu/m_{\rm PS} \lesssim 1.28$).

\section{Simulation results: Phase diagram at $T=0.89T_c$}
\label{sec:highTphase}
Next, we investigate  the phase diagram at $T=0.89T_c$. 
 This temperature 
is significantly higher than $T=0.45T_c$ as depicted in section 
\ref{sec:lowTphase} but still below $T_c$, the chiral transition temperature 
at zero chemical potential.   It is noteworthy that we can carry out the 
HMC simulation without the diquark source term even in the high chemical 
potential regime, $\mu/m_{\rm PS}\le 1.61~ (a\mu \le 1.00)$.  This suggests that at $T=0.89T_c$, superfluidity does not 
appear  at any density.

\begin{figure}[h]
\centering\includegraphics[width=15cm]{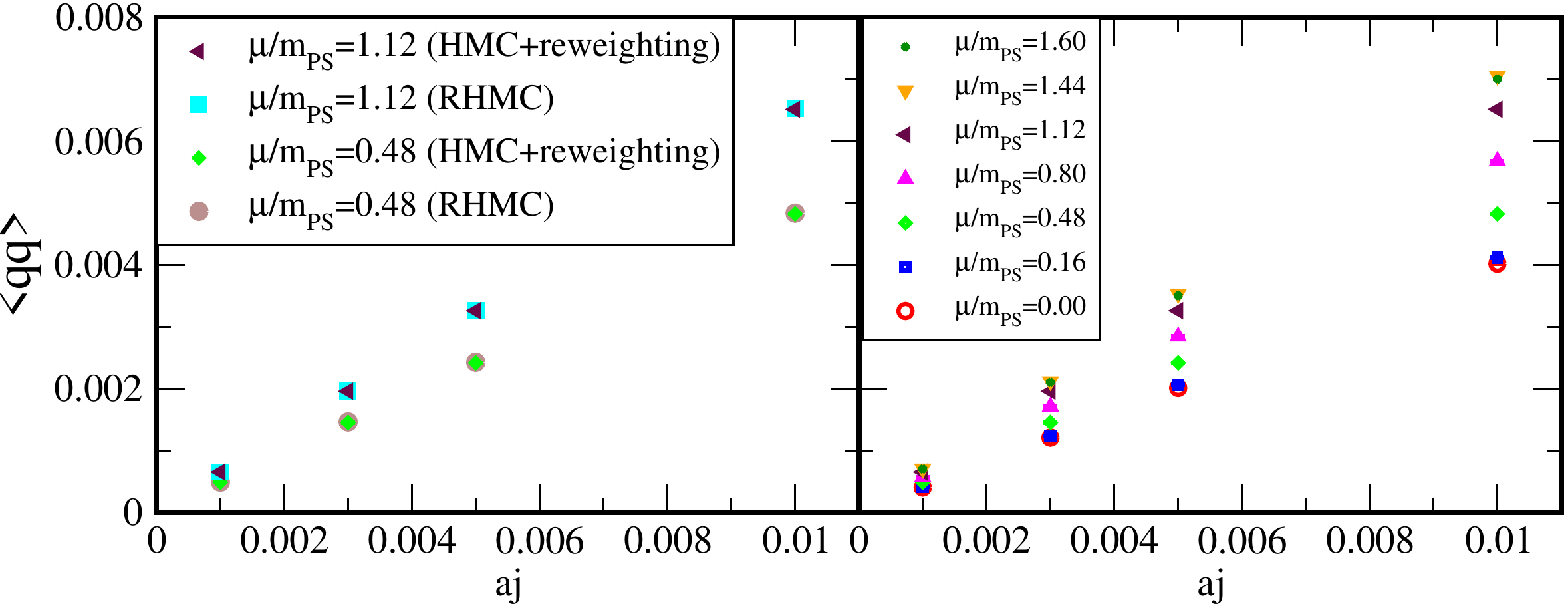}
\caption{The $j$-dependence of the diquark condensate at $T=0.89T_c$. The left panel shows the consistency between the data obtained by the RHMC and by the HMC with the reweighting method. The data in the right panel are obtained by the HMC with the reweighting method.}
\label{fig:diquark-Tc}
\end{figure}

To clarify the absence of superfluidity, we have performed the RHMC 
simulation with the diquark source term ($aj=0.001,0.003,0.005,$ and $0.01$) at $a\mu=0.30$ and $0.70$.
In the left panel of figure~\ref{fig:diquark-Tc}, there are two types of data for each $a\mu$: The one is obtained by the reweighting method for nonzero $aj$ where the configurations are generated by the HMC algorithm without the $j$ term. The other one is generated by the RHMC algorithm for each value of $aj$. 
These data are consistent within the statistical error bars. 
In the right panel of figure~\ref{fig:diquark-Tc}, we show the data obtained in the former way for several values of $\mu$ satisfying $a\mu \le 1.0$.    
By carrying out a linear extrapolation in terms of $aj$, we confirm that the diquark condensates at $j=0$ are consistent with zero.
Thus, the superfluidity does not appear at this temperature.  
%%%%%%%%%%%%%%%%%%
%%%%%%%%%%%%%%%%%%

\begin{figure}[h]
\centering\includegraphics[width=15cm]{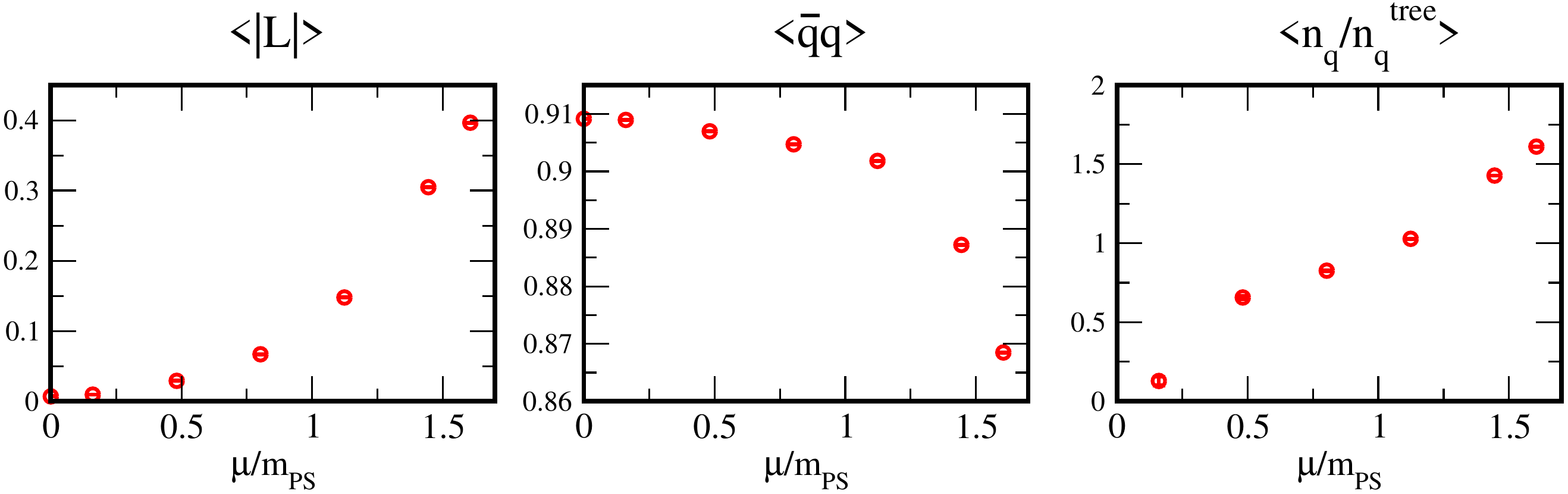}
\caption{The Polyakov loop (left), the chiral condensate (middle) and the 
quark number density (right) as a function of $\mu$.}
\label{fig:phase-Tc}
\end{figure}

By using the configurations obtained from the HMC simulation without the diquark source term in the lattice action, we have 
also measured the Polyakov loop, the chiral condensate, and the quark number density as shown in figure~\ref{fig:phase-Tc}.
%%%%%%%%%%%%%%%%%%
%%%%%%%%%%%%%%%%%%
The result for the Polyakov loop suggests that as $\mu$ 
increases, the system tends to be deconfined, while simultaneously the chiral 
symmetry tends to be restored.  Furthermore, the quark number density  has 
a small but nonzero value already around $\mu/m_{\rm PS}=0.16$, which is 
significantly smaller than the critical value of $\mu_B/m_{\rm PS}=1/2$, which is given by the ChPT and the lattice study at $T=0.45T_c$.
Note that as long as $\mu>0$, nonzero $n_q$ can occur via thermal 
excitations.  We can thus conclude that at $T=0.89T_c$, there is only a transition 
or crossover from the hadronic to the QGP phase at nonzero $\mu$ that is
closer to $\mu/m_{\rm PS}=0.16$ rather than $\mu/m_{\rm PS}=1/2$, 
while the superfluid phase does not appear even in the high density regime  below $T_c$.

\section{Simulation results: The topology at finite density}\label{sec:topo}
We are now in a position to examine the topological charge (or instanton 
number) in each of the phases as elucidated in sections \ref{sec:lowTphase} and 
\ref{sec:highTphase}.  The topological charge  can be measured by 
incorporating the gluonic definition,
\beq
Q = \frac{1}{32 \pi^2} \sum_{x} \mbox{Tr} \epsilon_{\mu \nu \rho \sigma} 
F^a_{\mu \nu} (x) F^a_{\rho \sigma} (x),
\eeq
into the gradient flow method.  
In the gradient flow process, we utilize the Wilson-plaquette action as a kernel of the flow equation. The field strength is calculated by using
the clover-leaf operator~\cite{Luscher:2010iy}. 
The number of configurations measured for each  set of the lattice parameters  has 
been set to $50$-$100$, while the flow time of the gradient flow has 
been fixed at $t/a^2=40$.

\begin{figure}[h]
\vspace{0.5cm}
\centering\includegraphics[width=14cm]{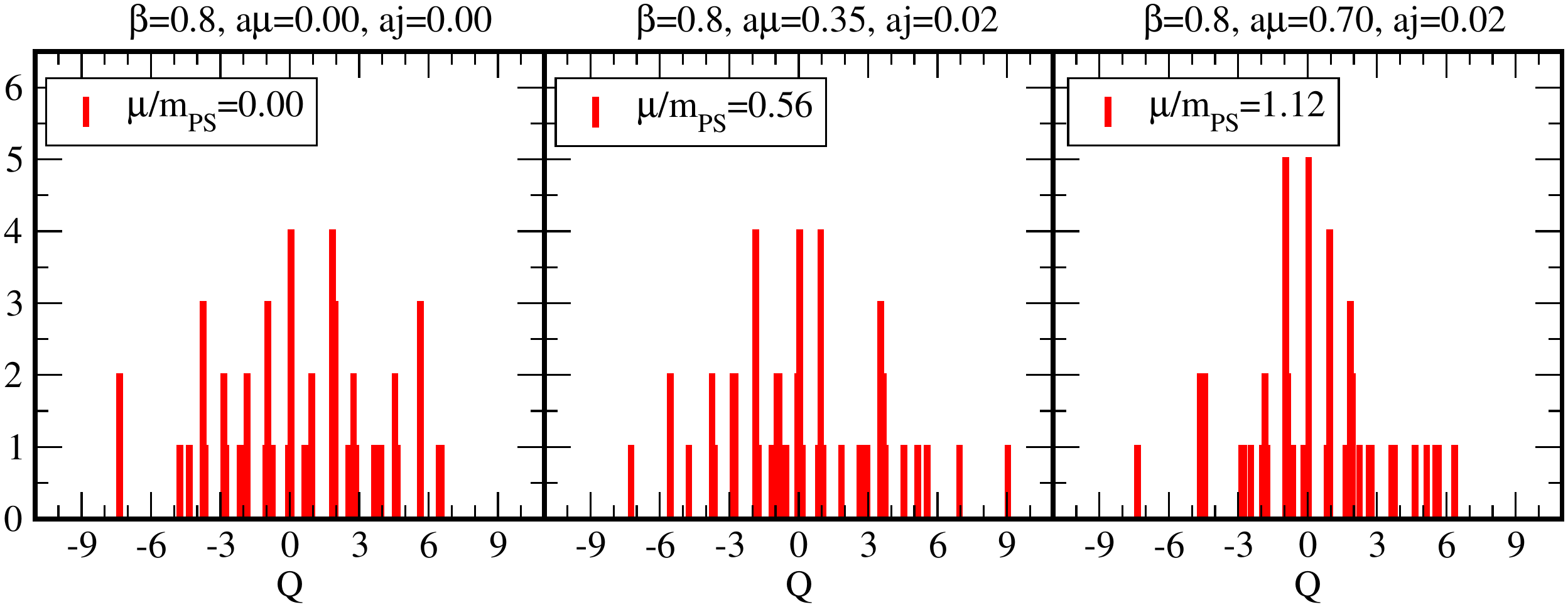}
\caption{Histogram of the topological charge at $\mu/m_{\rm PS}=0.00$ 
(left), 0.56 (middle) and 1.12 (right) in $\beta=0.800, N_s=N_\tau=16$ lattices.
The data are obtained from configurations generated with $aj=0.00$ for
$\mu/m_{\rm PS}=0.00$ and with $aj=0.02$ for $\mu/m_{\rm PS}=0.56, 1.12$.
}
\label{fig:hist-Q-beta08-Tc2}
\end{figure}

Figure \ref{fig:hist-Q-beta08-Tc2}  depicts the histogram of the 
topological charge  obtained at $T=0.45T_c$ for $\mu/m_{\rm PS}=0.00$, 
$0.56$ and $1.12$ ($a \mu=0.00$, $0.35$ and $0.70$), respectively.  Note 
that these three cases are typical of the hadronic, BEC and BCS phases, 
respectively.
%%%%%%%%%%%%%%%%%%
%%%%%%%%%%%%%%%%%%
Remarkably, no clear difference can be seen among them.  This result 
could have consequence to the diquark gap because the instanton density is 
expected to control the strength of the attraction between quarks 
\cite{Rapp:1997zu}.

This result is qualitatively different from what earlier investigations 
found for the SU(2) $N_f=4$ theory on $12^3 \times 24$ lattice 
\cite{Hands:2011hd} and SU($2$) $N_f=8$ theory on $14^3 \times 6$ lattice 
\cite{Alles:2006ea}.  In  such investigations, the topological 
susceptibility was found to decrease in the high chemical potential 
regime.  To understand such a qualitative difference, we also 
investigate the temperature and the phase dependences of the topological 
susceptibility.   

%%%%%%%%%%%%%%%%%%
\begin{figure}[h]
%\vspace{0.5cm}
\centering\includegraphics[width=14cm]{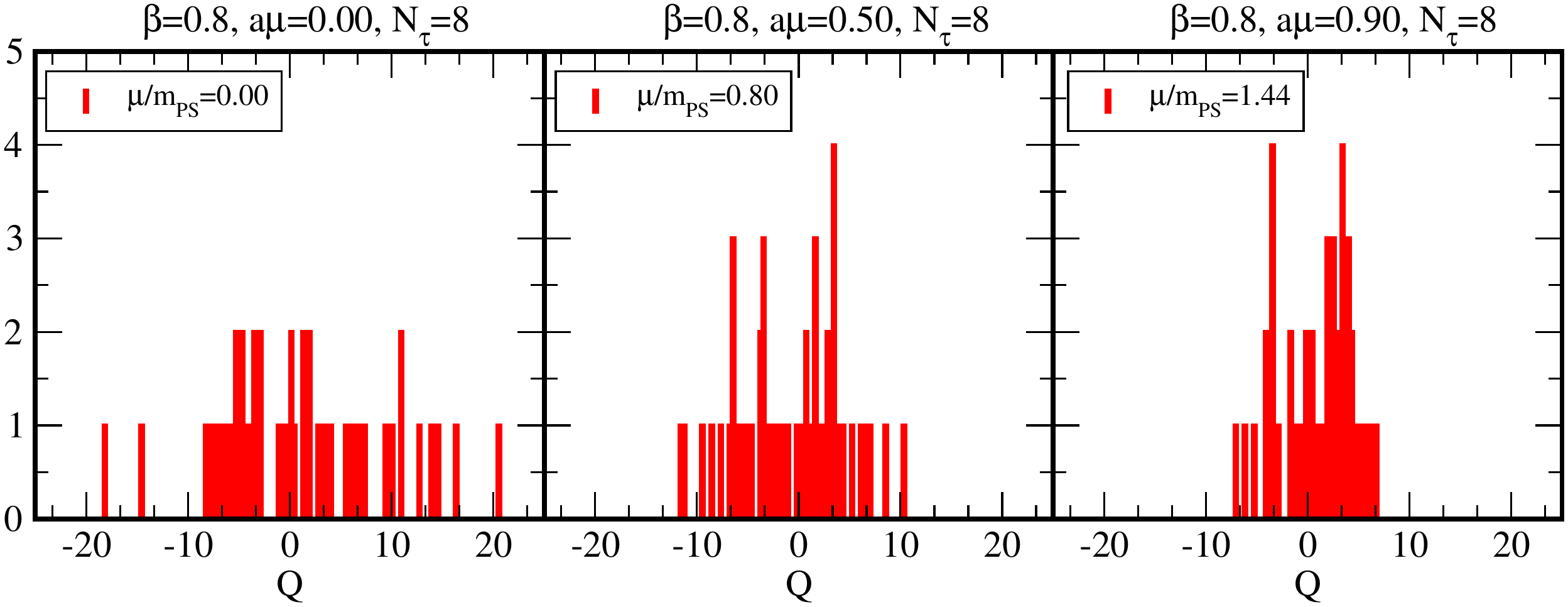}
\caption{Histogram of the topological charge  at $\mu/m_{\rm PS}=0.00$ 
(left), $0.80$ (middle) and $1.44$ (right) in $\beta=0.800, N_s=32, N_\tau=8$ 
lattices.   The data are obtained from configurations generated with 
$aj=0.00$.}
\label{fig:hist-Q-beta08-Tc}
%\vspace{6cm}
\end{figure}
%%%%%%%%%%%%%%%%%%

Figure~\ref{fig:hist-Q-beta08-Tc} shows the histogram of $Q$ obtained at 
$T=0.89T_c$, where the phase changes from the hadronic to QGP one with increasing $\mu$.  At zero $\mu$, 
there is still a broad distribution of $Q$, while in the high chemical 
potential regime ($a\mu=0.90$), the distribution of $Q$ becomes narrow.  Consequently, the topological susceptibility decreases with 
$\mu$.

\begin{figure}[h]
%\vspace{0.5cm}
\centering\includegraphics[width=14cm]{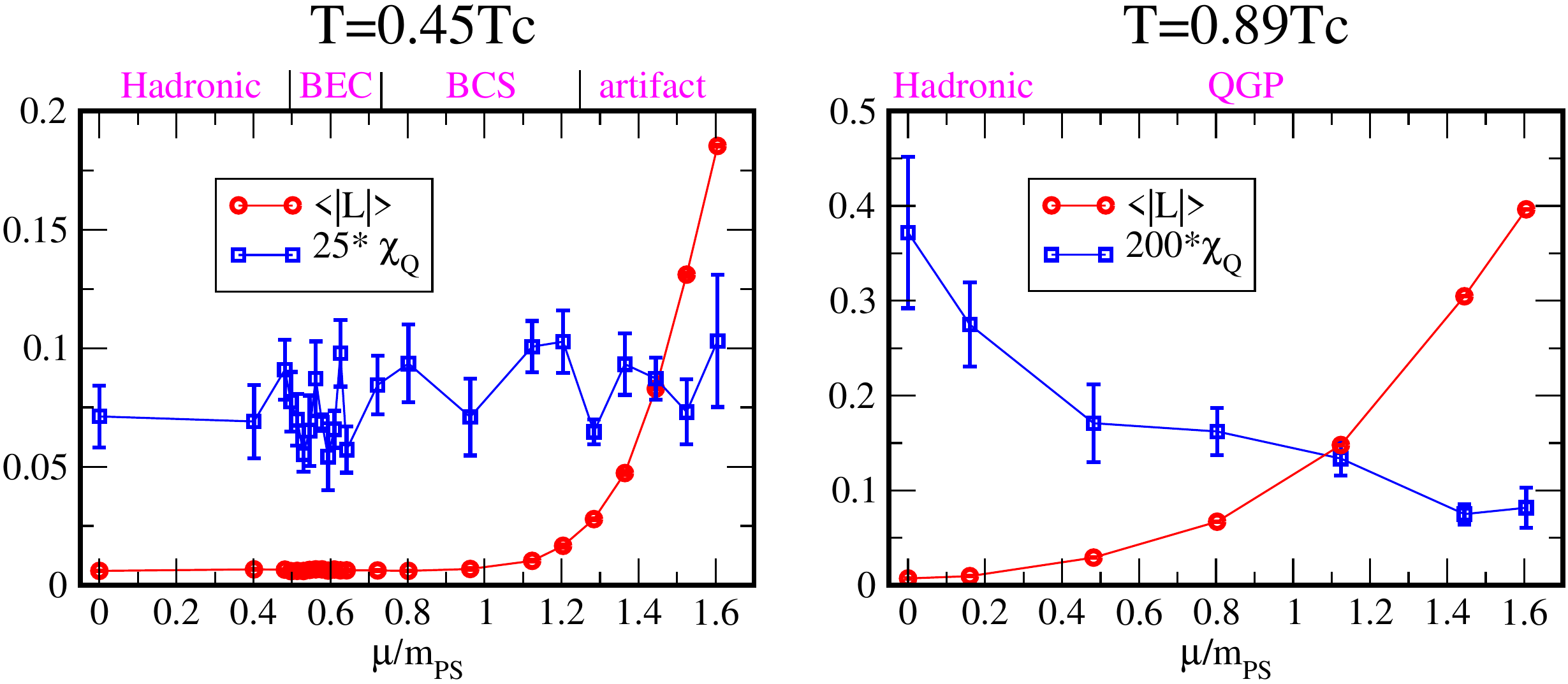}
\caption{Polyakov loop and topological susceptibility in 
$\beta=0.800, N_s=N_\tau=16$ (left) and $\beta=0.800, N_s=32, N_\tau=8$ (right) 
lattices, plotted as a function of $\mu$.  The data for 
$\mu /m_{\rm PS}>0.50$ in the left panel are  obtained from the 
configurations generated with $aj=0.02$, while the others are obtained with 
$aj=0.00$.}
\label{fig:Ploop-Topology}
%\vspace{6cm}
\end{figure}

We conclude this section by showing, in figure \ref{fig:Ploop-Topology}, 
how the topological susceptibility and the Polyakov loop depend on $\mu$ at 
$T=0.45T_c$ and $0.89T_c$.  
Here, to make it easy to see,  $\chi_Q$ multiplied by a constant $25$ and $200$ are shown in the left and right panels, respectively.
As can be already seen from figures 
\ref{fig:hist-Q-beta08-Tc2} and \ref{fig:hist-Q-beta08-Tc}, there is a crucial 
difference between these two cases: At $T=0.89T_c$ the topological 
susceptibility decreases with $\mu$, which seems just opposite to the behaviour 
of the Polyakov loop, whereas at $T=0.45T_c$ the topological susceptibility is 
almost independent of $\mu$ even in the regime where the Polyakov loop
significantly increases.
The qualitative difference comes from the underlying phase structure.
%%%%%%%%%%%%%%%%%%
%%%%%%%%%%%%%%%%%%

\section{Outlook}
\label{sec:outlook}
In this paper we have laid out the phase and topological structure of 
dense two-colour QCD on lattice.  In addition to accuracy improvement, however, 
many questions remain.  
First, scale setting \cite{Lee} has to be finalized for
a continuum extrapolation to remove lattice artifacts.  
Second, it would 
be significant to find the critical temperature at which the diquark condensate 
disappears in high density regime.  
Especially in the BCS phase, such critical temperature would lead 
to estimates of the zero-temperature diquark gap given in eq.~(\ref{eq:BCS-relation}).  
Finally, it would be 
interesting to extend the present work to the cases of lower temperature 
and larger lattice.  
These lattice studies would help us study 
the hadron masses~\cite{Muroya:2002ry, Hands:2007uc,Wilhelm:2019fvp} and the hadron-hadron interactions~\cite{Ishii:2006ec,Aoki:2009ji,Takahashi:2009ef,Ikeda:2011bs,Amato:2015gea}.

%\appendix
%\section{Corresponding table of the phase in previous works}\label{sec:definition-phase}
%Please always give a title also for appendices.

\acknowledgments
We are grateful to  S.~Aoki, P.~de Forcrand, K.~Fukushima, K.~Ishiguro,  
K.~Nagata, A.~Nakamura, A.~Ohnishi and N.~Yamamoto for useful comments.
Discussions during the Yukawa Institute for Theoretical Physics (YITP)
long-term international workshop on "New Frontiers in QCD 2018" were
useful to initiate this work.
The numerical simulations were carried out on SX-ACE and OCTOPUS
 at Cybermedia Center (CMC) and Research Center for Nuclear Physics (RCNP), 
Osaka University, together with XC40 at YITP and Institute for Information Management and Communication 
(IIMC), Kyoto University.
We also acknowledge the help of CMC in tuning the gradient-flow code.
T.-G.~L. acknowledges the support of the Seiwa Memorial Foundation.
This work partially used computational resources of
 HPCI-JHPCN System Research Project (Project ID: jh180042 and jh190029) in Japan.
 This work was supported in part by Grants-in-Aid for Scientific Research 
through Grant Nos.
 15H05855,%(Nitta-san)
18H01217, 18H05406, 18H01211 %(Iida)
and 19K03875%(Itou)
, which were 
provided by the Japan Society for the Promotion of Science (JSPS), and in part 
by the Program for the Strategic Research Foundation at Private Universities 
``Topological Science'' through Grant No.\ S1511006, which was supported by the
Ministry of Education, Culture, Sports, Science and Technology (MEXT) of 
Japan.

\end{document}